\documentclass{llncs}

\usepackage{graphicx}

\usepackage{booktabs}
\usepackage{tabularx}

\usepackage{subcaption}
\usepackage{subfig}

\title{Characterizing Scales of Genetic Recombination and Antibiotic Resistance in Pathogenic Bacteria Using Topological Data Analysis}

\author{Kevin J. Emmett\inst{1} \and Raul Rabadan\inst{2}}

\institute{
Department of Physics, Columbia University \email{kje2109@columbia.edu}
\and
Department of Systems Biology and Department of Biomedical Informatics, Columbia University \email{rr2579@c2b2.columbia.edu}
}

\begin{document}

\maketitle

\begin{abstract}
Pathogenic bacteria present a large disease burden on human health.
Control of these pathogens is hampered by rampant lateral gene transfer, whereby pathogenic strains may acquire genes conferring resistance to common antibiotics.
Here we introduce tools from topological data analysis to characterize the frequency and scale of lateral gene transfer in bacteria, focusing on a set of pathogens of significant public health relevance.
As a case study, we examine the spread of antibiotic resistance in \emph{Staphylococcus aureus}.
Finally, we consider the possible role of the human microbiome as a reservoir for antibiotic resistance genes.

\keywords{topological data analysis, microbial evolution, antibiotic resistance}
\end{abstract}

\section{Introduction}
\label{sec:introduction}

Pathogenic bacteria can lead to severe infection and mortality and presents an enormous burden on human populations and public health systems.
One of the achievements of twentieth century medicine was the development of a wide range of antibiotic drugs to control and contain the spread of pathogenic bacteria, leading to vastly increased life expectancies and global economic development.
However, rapidly rising levels of multidrug antibiotic resistance in several common pathogens, including \emph{Escherichia coli}, \emph{Klebsiella pneumoniae}, \emph{Staphylococcus aureus}, and \emph{Neisseria gonorrhoea}, is recognized as a pressing global issue with near-term consequences \cite{Neu:1992gk,Thomas:2005hp,WHO:2014wa}.
The threat of a post-antibiotic 21st century is serious, and new methods to characterize and monitor the spread of resistance are urgently needed.

Antibiotic resistance can be acquired through point mutation or through horizontal transfer of resistance genes.
Horizontal exchange occurs when a donor bacteria transmits foreign DNA into a genetically distinct bacteria strain.
Three mechanisms of horizontal transfer are identified, depending on the route by which foreign DNA is acquired \cite{Ochman:2000dr}.
Foreign DNA can be acquired via uptake from an external environment (transformation), via viral-mediated processes (transduction), or via direct cell-to-cell contact between bacterial strains (conjugation).
Resistance genes can be transferred between strains of the same species, or can be acquired from different species in the same environment.
While the former is generally more common, an example of the latter is the phage-mediated acquisition of Shiga toxin in \emph{E. coli} in Germany in 2011 \cite{Rohde:2011ju}.
Elements of the bacterial genome that show evidence of foreign origin are called genomic islands, and are of particular concern when associated with phenotypic effects such as virulence or antibiotic resistance.

The presence of horizontal gene transfer precludes accurate phylogenetic characterization, because different segments of the genome will have different evolutionary histories.
Bacterial species definitions and taxonomic classifications are made on the basis of 16S ribosomal RNA, a highly conserved genomic region between bacteria and archaea species \cite{Woese:1977vd}.
However, the region generally accounts for less than 1\% of the complete genome, implying that the vast majority of evolutionary relationships are not accounted for in the taxonomy \cite{Dagan:2006up}.
Because of the important role played by lateral gene transfer, new ways of characterizing evolutionary and phenotypic relationships between microorganisms are needed.

Topological data analysis (TDA), and persistent homology in particular, has been shown to be an effective tool for capturing horizontal evolutionary processes at the population level by measuring deviations from treelike additivity.
Initial work in this direction characterized recombination in viral evolution, particularly in influenza, where genomic reassortment can lead to the emergence of viral pandemics \cite{Chan:2013vt}.
Further work established foundations for statistical inference in population genetics models using TDA \cite{Emmett:2014xx}.
We provide a brief overview of TDA and persistent homology in Section \ref{section:tda}, for additional reviews see \cite{Carlsson:2009vh}\cite{Edelsbrunner:2010vl}.

In this paper we explore topics relating to horizontal gene transfer in bacteria and the emergence of antibiotic resistance in pathogenic strains.
We show that TDA can not only quantify gene transfer events, but also characterize the scale of gene transfer.
The scale of recombination can be measured from the distribution of birth times of the $H_1$ invariants in the persistent barcode diagram.
It has been shown that recombination rates decrease with increasing sequence divergence \cite{Fraser:2007ep}.
We characterize the rate and scale of intraspecies recombination in several pathogenic bacteria of public health concern.
We select a set of pathogenic bacteria that are of significant public health interest based on a recently released World Health Organization (WHO) report on antimicrobial resistance \cite{WHO:2014wa}.
Using persistent homology, we characterize the rate and scale of recombination in the core genome using multilocus sequence data.
To extend our characterization to the whole genome, we use protein family annotations as a proxy for sequence composition.
This allows us to compute a similarity matrix between strains.
Comparing persistence diagrams gives us information about the relative scales of gene transfer at arbitrary loci.
The species selected for study and the sample sizes in each analysis are specified in Table~\ref{table:samplesizes}.
Next, we explore the spread of antibiotic resistance genes in \emph{S. aureus} using Mapper, an algorithm for partial clustering and visualization of high dimensional data \cite{Singh:2007ve}.
We identify two major populations of \emph{S. aureus}, and observe one cluster with strong enrichment for the antibiotic resistance gene \emph{mecA}.
Importantly, resistance appears to be increasingly spreading in the second population.
Finally, we consider the risk of horizontal transfer of resistance genes from the human microbiome into an antibiotic sensitive strain, using $\beta$-Lactam resistence as an example.
In this environment, benign bacterial strains can harbor known resistance genes.
We use a network analysis to visualize the spread of antibiotic resistance gene $\emph{mecA}$ into nonnative phyla.
Each individual has a unique microbiome, and we speculate that microbiome typing of this sort may useful in developing personalized antibiotic therapies.

These results demonstrate the important role the HCI-KDD approach can play in tackling the challenges of large scale -omics data applied to clinical settings and personalized medicine:
Interactive visualization through graph and network construction, data mining global invariants with topological algorithms, and knowledge discovery through data integration and fusion \cite{Holzinger:2013xx}\cite{Holzinger:2014vz}.

\begin{table}[t]
\centering
\caption{Pathogenic bacteria selected for study and sample sizes in each analysis.}
\begin{tabularx}{\textwidth}{Xcc}
\toprule
Species & \hspace{10mm}MLST profiles\hspace{10mm} & PATRIC profiles \\
\midrule
\emph{Campylobacter jejuni}      & 7216  & 91 \\
\emph{Escherichia coli} 		 & 616   & 1621 \\
\emph{Enterococcus faecalis} 	 & 532   & 301 \\
\emph{Haemphilus influenzae} 	 & 1354  & 22 \\
\emph{Helicobacter pylori} 	     & 2759  & 366 \\
\emph{Klebsiella pneumoniae}     & 1579  & 161 \\
\emph{Neisseria spp.}       	 & 10802 & 234 \\
\emph{Pseudomonas aeruginosa} 	 & 1757  & 181 \\
\emph{Staphylococcus aureus}     & 2650  & 461 \\
\emph{Salmonella enterica}   	 & 1716  & 638 \\
\emph{Streptococcus pneumoniae}  & 9626  & 293 \\
\emph{Streptococcus pyogenes}    & 627   & 48 \\
\bottomrule
\end{tabularx}
\label{table:samplesizes}
\end{table}

\section{Topological Data Analysis and Persistent Homology}
\label{section:tda}
Topological data analysis computes global invariants from point cloud data.
These global invariants represent loops, holes, and higher dimensional voids in data.
A topological representation of the data is constructed by building a set of triangulated objects representing the connectivity of the data at different scales, called a filtration.
Various constructions exist for triangulating data.
The most efficient approach for large scale data is the Vietoris-Rips complex, which associates a simplex to a set of points if they are pairwise connected.
In this way, the complex is specified purely by its 1-skeleton, which can be efficiently computed.

Given a filtration, persistent homology is an algorithm to associate homology groups to each scale, which give information about the invariants in the data.
$H_0$ gives information about the connectivity, $H_1$ about loops, etc.
The output of the algorithm is a set of intervals corresponding to topological features present at different scales.
The homology information can be compactly summarized as a barcode diagram, in which invariants are represented as horizontal line segments with the birth and death scales as the left and right edge of the line segment, respectively.
Alternatively, the homology information can be represented as a persistence diagram, a 2-D plot in which intervals are represented as points on a (birth, death) plane.

For the purposes of studying biological sequence data, and horizontal evolutionary processes in particular, these approaches are widely useful, as it was shown in \cite{Chan:2013vt} that sequence datasets with treelike phylogeny will have vanishing higher homology.
The observed homological features are therefore direct evidence for horizontal exchange amonst the sequences in a sample.
These topological constructions become a natural way of reasoning about evolutionary relationships between organisms in cases where treelike phylogeny is not appopriate.

\section{Evolutionary Scales of Recombination in the Core Genome}
\label{sec1}
The genes comprising the bacterial genome can be largely paritioned into two groups: the core genome, consisting of those genes that are highly conserved and characteristic of a given species classification, and the accessory genome, consisting of those genes whose presence can be variable even within strains of the same species.
We first sought to examine scales of recombination in the core bacterial genome using multilocus sequence typing (MLST) data.
MLST is a method of rapidly assigning a sequence profile to a sample bacterial strain.
For each species, a predetermined set of loci on a small number of housekeeping genes are selected as representative of the core genome of the species.
As new strains are sequenced, they are annotated with a profile corresponding to the sequence type at each locus.
If a sample has a previously unseen type at a given locus, it is appended to the list of types at that locus.
Large online databases have curated MLST data from labs around the world; significant pathogens can have several thousand typed strains (over 10,000 in the case of \emph{Neisseria spp.}).
Because different species will be typed at different loci, examining direct interspecies genetic exchange with this data is unfeasible, however MLST provides a large quantity of data with which to examine intraspecies exchange in the core genome.
However, because the selected loci are generally all housekeeping genes, this type of recombination analysis will be only informative about genetic exchange in the core genome.
Mobile genetic elements will have separate rates of exchange.

\begin{figure}
\centering
\begin{subfigure}{0.48\textwidth}
\centering
\includegraphics[width=\textwidth]{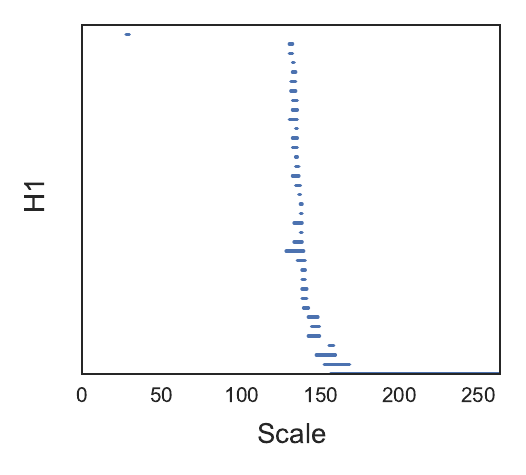}
\caption{\emph{Klebsiella pneumoniae}}
\end{subfigure}
\begin{subfigure}{0.48\textwidth}
\centering
\includegraphics[width=\textwidth]{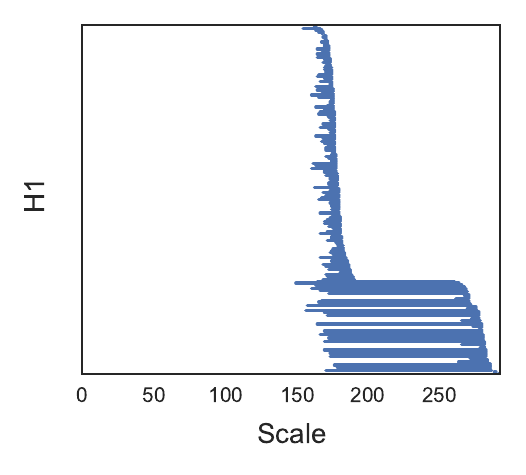}
\caption{\emph{Salmonella enterica}}
\end{subfigure}
\caption{Barcode diagrams reflect different scales of genomic exchange in \emph{K. pneumoniae} and \emph{S. enterica}.}
\label{fig:fig1_barcode}
\end{figure}

\begin{figure}
\centering
\includegraphics[width=\textwidth]{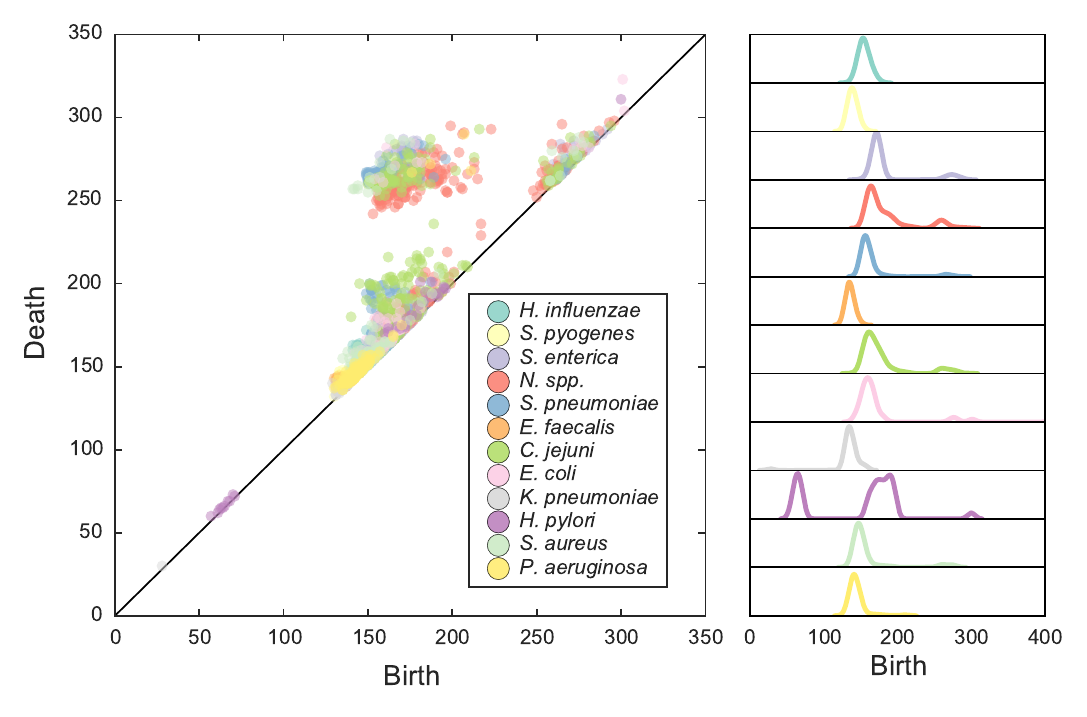}
\caption{The $H_1$ persistence diagram for the twelve pathogenic strains selected for this study using MLST profile data. There are three broad scales of recombination. To the right is the birth time distribution for each strain. \emph{H. pylori} has an earlier scale of recombination not present in the other species.}
\label{fig2:mlst_persistence_diagram}
\end{figure}

We investigate horizontal exchange in the core genome for twelve pathogens using MLST data from PubMLST \cite{Jolley:2010gf}.
For each strain, a pseudogenome can be constructed by concatenating the typed sequence at each locus.
Using a Hamming metric, we construct a pairwise distance matrix between strains and compute persistent homology on the resulting metric space.
Because of the large number of sample strains, we employ a Lazy Witness complex with 250 landmark points and $\nu=0$ \cite{deSilva:2004tg}.
The computation is performed using javaplex \cite{javaplex:2011xx}.
An example of our output is shown in Figure~\ref{fig:fig1_barcode}, where we plot the $H_1$ barcode diagrams for \emph{K. pneumoniae} and \emph{S. enterica}.
The two species have distinct recombination profiles, characterized by the range of recombinations: \emph{K. pneumoniae} recombines at only one short-lived scale, while \emph{S. enterica} recombines both at the short-lived scale and a longer-lived scale.
We repeat this analysis for each species, and plot the results as a persistence diagram in Figure~\ref{fig2:mlst_persistence_diagram}.
Among the bulk of pathogens there appears to be three major scales of recombination, a short-lived scale at intermediate distances, a longer-lived scale at intermediate distances, and a short-lived scale at longer distances.
\emph{H. polyori} is a clear outlier, tending to recombine at scales significantly lower than the other pathogens.

\begin{figure}
\centering
\includegraphics[width=\textwidth]{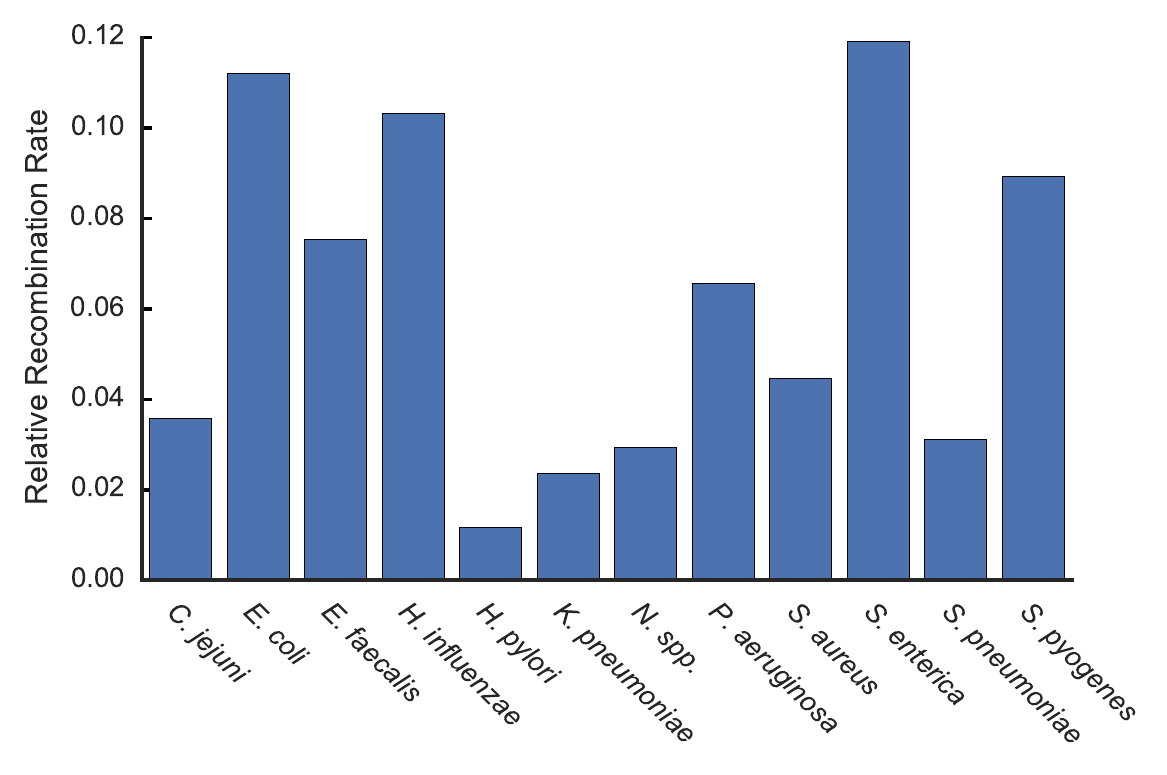}
\caption{Relative recombination rates computed by persistent homology from MLST profile data.}
\label{fig:barchart}
\end{figure}

We define a relative rate of recombination by counting the number of $H_1$ loops across the filtration and dividing by the number of samples for that species.
The results are shown in Figure~\ref{fig:barchart}, where we observe that different species can have vastly different recombination profiles.
For example, \emph{S. enterica} and \emph{E. coli} have the highest recombination rates, while \emph{H. pylori} is substantially lower than the others.
Coupled with the smaller scale of recombinations suggests that the \emph{H. pylori} core genome is relatively resistant to recombination except within closely related strains.

\section{Protein Families as a Proxy for Genome Wide Reticulation}
\label{sec2}

Protein family annotations cluster proteins into sets of isofunctional homologs, i.e., clusters of proteins with both similar sequence composition and similar function.
A particular strain is represented as a binary vector indicating the presence or absence of a given protein family.
Correlations between strains can reveal genome-wide patterns of genetic exchange, unlike the MLST data which can only provide evidence of exchange in the core genome.
We use the FigFam protein annotations in the Pathosystems Resource Institute Center (PATRIC) database because of the breadth of pathogenic strain coverage and depth of genomic annotations \cite{Wattam:2013jy}.
The FigFam annotation scheme consists of over 100,000 protein families curated from over 950,000 unique proteins \cite{Meyer:2009iq}.

\begin{figure}
\centering
\includegraphics[width=\textwidth]{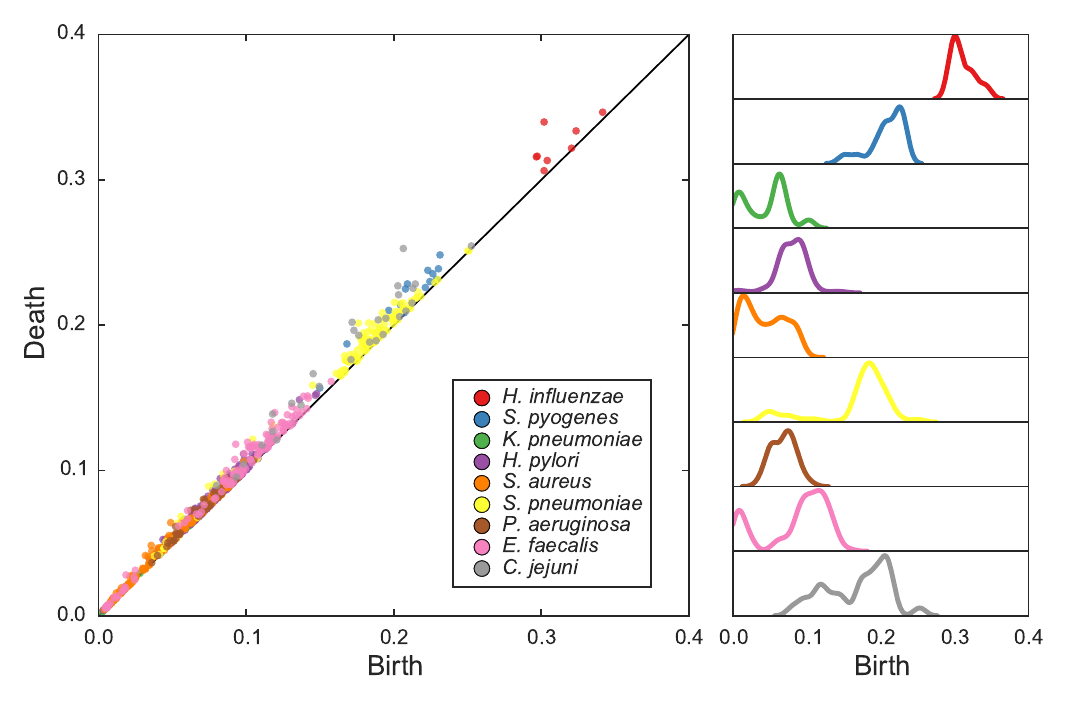}
\caption{Persistence diagram for a subset of pathogenic bacteria, computed using the FigFam annotations compiled in PATRIC. Compared to the MLST persistence diagram, the Figfam diagram has a more diverse scale of topological structure.}
\label{fig:figfam_persistence_diagram}
\end{figure}

For each strain we compute a transformation into FigFam space.
We transform into this space because the frequency of genome rearrangements and differences in mobile genetic elements makes whole genome alignments unreliable, even for strains within the same species.
As justification for performing this step, it has been shown experimentally that recombination rates decrease with increasing genetic distance \cite{Fraser:2007ep}.
After transforming, we construct a strain-strain correlation matrix and compute the persistent homology in this space.
In Figure~\ref{fig:figfam_persistence_diagram} we show the persistence diagram relating the structure and scale between different species.
We find that different species have a much more diverse topological structure in this space than in MLST space, and a wide variety of recombination scales.
The large scales of exchange in \emph{H. influenzae} suggest it can regularly acquire novel genetic material from distantly related strains.

\section{Antibiotic Resistance in \emph{Staphylococcus aureus}}

\emph{S. aureus} is a gram positive bacteria commonly found in the nostrils and upper respiratory tract.
Certain strains can cause severe infection in high-risk populations, particulary in the hospital setting.
The emergence of antibiotic resistant \emph{S. aureus} is therefore of significant clinical concern.
Methicillin resistant \emph{S. aureus} (MRSA) strains are resistant to $\beta$-Lactam antibiotics including penicillin and cephalosporin.
Resistance is conferred by the gene \emph{mecA}, an element of the Staphyloccoccal cassette chromosome mec (\emph{SCCmec}).
\emph{mecA} codes for a dysfunctional penicillin-binding protein 2a (PBP2a), which inhibits $\beta$-Lactam antibiotic binding, the primary mechanism of action \cite{Jensen:2009fu}.
Of substantial clinical importance are methods for characterizing the spread of MRSA within the \emph{S. aureus} population.

\begin{figure}[t]
\centering
\includegraphics[width=\textwidth]{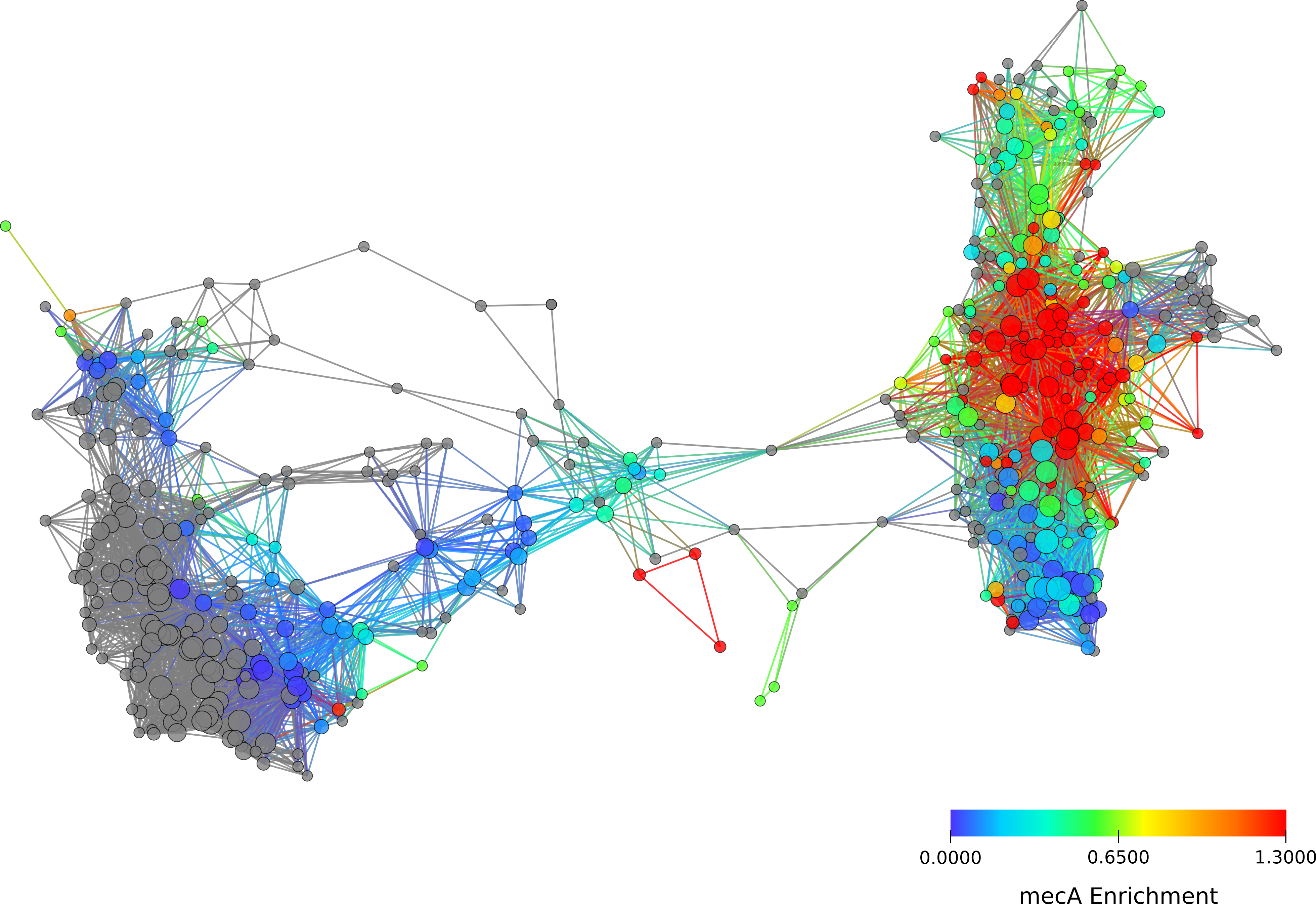}
\caption{The FigFam similarity network of \emph{S. aureus} constructed using Mapper as implemented in Ayasdi Iris. We use a Hamming metric and Primary and Secondary Metric SVD filters (res: 30, gain 4x, eq.). Node color is based on strain enrichment for \emph{mecA}, the gene conferring $\beta$-Lactam resistance. Two distinct clades of \emph{S. aureus} are visible, one of which has already been compromised for resistance. Of important clinical significance is the growing enrichment for \emph{mecA} in the second clade.}
\label{fig:saureus_network}
\end{figure}

To address this question, we use the FigFam annotations in PATRIC, as described in the previous section.
PATRIC contains genomic annotations for 461 strains of \emph{S. aureus}, collectively spanning 3,578 protein families.
We perform a clustering analysis using the Mapper algorithm as implemented in Ayasdi Iris \cite{AyasdiIris}.
Principal and second metric singular value decomposition are used as filter functions, with a 4x gain and an equalized resolution of 30.
This results in a graph structure with two large clusters, connected by a narrow bridge, as shown in Figure~\ref{fig:saureus_network}.
The two clusters are consistent with previous phylogenetic studies using multilocus sequence data to identify two major population groups \cite{Cooper:2006dp}.

Of the 461 \emph{S. aureus} strains in PATRIC, 142 carry the \emph{mecA} gene.
When we color nodes in the network based on an enrichment for the presence of \emph{mecA}, we observe a much stronger enrichment in one of the two clusters.
This suggests that $\beta$-Lactam resistance has already begun to dominate in that clade, likely due to selective pressures.
More strikingly, we observe that while \emph{mecA} enrichment is not as strong in the second cluster, there is a distinct path of enrichment emanating along the connecting bridge between the two clusters and into the less enriched cluster.
This suggests the hypothesis that antibiotic resistance has spread from the first cluster into the second cluster via strains intermediate to the two, and will likely continue to be selected for in the second cluster.

\section{Microbiome as a Reservoir of Antibiotic Resistance Genes}

While antibiotic resistance can be acquired through gene exchange between strains of the same species, it is also possible for gene exchange to occur between distantly related species.
It has been recognized that an individual's microbiome, the set of microorganisms that exist symbiotically within a human host, can act as a reservoir of antimicrobial resistance genes \cite{Sommer:2010uh,Penders:2013wt}.
It is of substantial clinical interest to characterize to what extent an individual's microbiome may pose a risk for a pathogenic bacteria acquiring a resistance gene through horizontal transfer from an benign strain in the microbiome.

\begin{figure}[t]
\centering
\includegraphics[width=\textwidth]{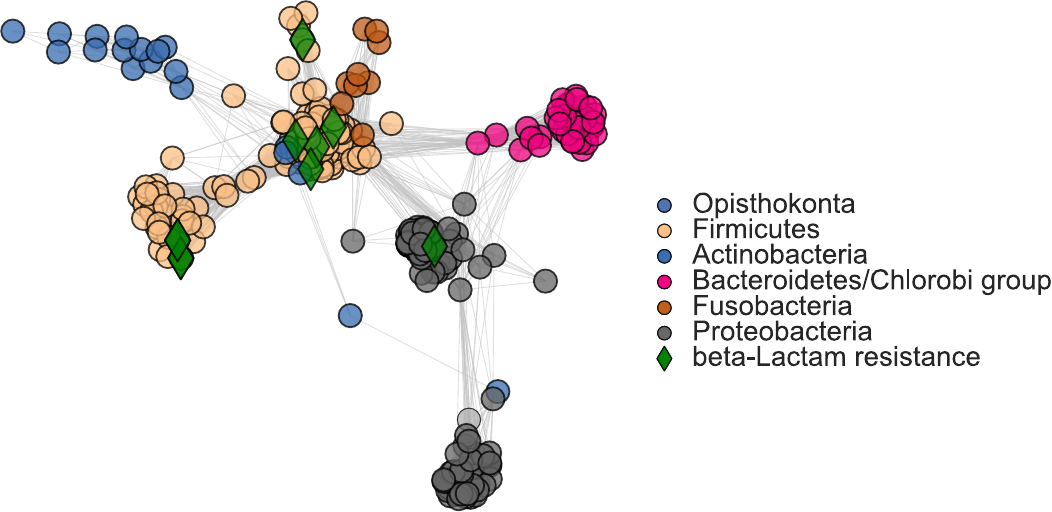}
\caption{The FigFam similarity network of gastrointestinal tract reference strains identified in the Human Microbiome Project. The green diamond identifies the strains carrying resistance to $\beta$-Lactam antibiotics.}
\label{fig:microbiome_network}
\end{figure}

To address this question, we use data from the Human Microbiome Project (HMP), a major research initiative performing metagenomic characterization of hundreds of healthy human microbiomes \cite{Consortium:2012bm}.
The HMP has defined a set of reference strains that have been observed in the human microbiome.
We collect FigFam annotations from PATRIC for the reference strain list in the gastrointestinal tract.
We focus on the gastrointestinal tract because it is an isolated environment and likely to undergo higher rates of exchange than other anatomic regions.
Of the 717 gastrointestinal tract reference strains, 321 had FigFam annotations.
We computed a similarity matrix as in previous sections, using correlation as distance.
The resulting network is shown in Figure~\ref{fig:microbiome_network}, where strains are colored by phyla-level classifications.
While largely recapitulating phylogeny, the network depicts interesting correlations between phyla, such as the loop between Firmicutes, Bacteroides, and Proteobacteria.

Next, we searched for genomic annotations relating to $\beta$-Lactam resistance.
10 strains in the reference set had matching annotations, and we highlight those strains in the network with green diamonds.
We observe resistance mostly concentrated in the Firmicutes, of which \emph{S. aureus} is a member, however there is a strain of Proteobacteria that has acquired the resistance gene.
Transfer of $\beta$-Lactam resistance into the Protebacteria is clinically worrisome.
Pathogenic Proteobacteria include \emph{S. enterica}, \emph{V. cholerae}, and \emph{H. pylori}, and emergence of $\beta$-Lactam resistance will severely impact currently used antibiotic drug therapies.

The species composition of each individual's microbiome can differ substantially due to a wide variety of poorly understood factors \cite{Consortium:2012bm}.
In this case, an individual's personal microbiome network will differ from the network we show in Figure~\ref{fig:microbiome_network}, which was constructed from the set of \emph{all} strains that have been reported across studies of multiple individuals.
The relative risk for acquiring self-induced resistance will therefore vary from person to person and by the infectious strain acquired.
However, a network analysis of this type can be used to assess risk and give clues as to possible routes by which antibiotic resistance may be acquired.
In the clinical setting, this could assist in developing personalized antibiotic treatment regimens.
We propose a more thorough expansion of this work, examining the full range of antibiotic resistance genes in order to quantify microbiome risk factors for treatment failure.
We foresee an era of genomically informed infectious disease management in the clinical setting, based on an understanding of a patient's personal microbiome network profile.

\section{Conclusions}

In this paper we have brought some ideas from topological data analysis to bear on problems in pathogenic microbial genetics.
First, we used persistent homology to evaluate recombination rates in the bacterial core genomes using MLST profile data.
We showed that different pathogens have different recombination rates.
We expanded this to gene transfer across whole bacterial genomes by using protein family annotations in the PATRIC database.
We found different scales of recombination in different pathogens.
Next, we explored the spread of MRSA in \emph{S. aureus} populations using topological methods.
We identified two major population clusters of \emph{S. aureus}, and noted increasing resistance in a previously isolated population.
Finally, we studied the emergence of $\beta$-Lactam resistance in the microbiome, and proposed methods by which personal risk could be assessed by microbiome typing.
Each stage of this analysis represents a successful application of the HCI-KDD approach to biomedical discovery.
Our results point to the important role for graph mining and topological data mining in health and personalized medicine.

\subsubsection*{Acknowledgements}
The authors thank Gunnar Carlsson for access to the Ayasdi Iris platform.
KJE thanks Chris Wiggins, Daniel Rosenbloom, and Sakellarios Zairis for useful discussions.
KJE and RR were supported by NIH grant U54-CA121852, Multiscale Analysis of Genomic and Cellular Networks.

This publication made use of the PubMLST website (http://pubmlst.org/) developed by Keith Jolley \cite{Jolley:2010gf} and sited at the University of Oxford. The development of that website was funded by the Wellcome Trust.

\bibliographystyle{splncs03}
\bibliography{warsaw}

\end{document}